\documentclass[12pt]{article}

\usepackage{graphicx}
\usepackage{scicite}

\usepackage{times}

\newenvironment{sciabstract}{%
\begin{quote} \bf}
{\end{quote}}

\title{Microrobots from Toposelective Nanoparticle Attachment}

\author
{M. R. Bailey,$^{1\ast}$ F. Grillo,$^{1\ast}$ N. D. Spencer$^{2}$, L. Isa$^{1\ast}$\\

\\
\normalsize{$^{1}$Laboratory for Soft Materials and Interfaces, Department of Materials, ETH Z{\"u}rich,} \\ \normalsize{Vladimir-Prelog-Weg 5, 8093 Z{\"u}rich, Switzerland}\\
\normalsize{$^{2}$Laboratory for Surface Science and Technology, Department of Materials, ETH Z{\"u}rich,}\\
\normalsize{Vladimir-Prelog-Weg 5, 8093 Z{\"u}rich, Switzerland}
\\
\normalsize{$^\ast$To whom correspondence should be addressed:} \\
\normalsize{E-mail: maximilian.bailey@mat.ethz.ch;}\\ 
\normalsize{E-mail: fabio.grillo@alumni.ethz.ch;}\\ 
\normalsize{E-mail: lucio.isa@mat.ethz.ch}
}

\date{}

\begin{document} 


\baselineskip24pt


\maketitle


\begin{sciabstract}
Microrobots hold promise for applications ranging from targeted delivery to enhanced mixing at the microscale. However, current fabrication techniques suffer from limited throughput and material selection. Here, we demonstrate a versatile route enabling the synthesis of microrobots from off-the-shelf micro- and nano-particles. Our protocol hinges on the toposelective attachment of photocatalytic nanoparticles onto microparticles, exploiting a multi-functional polymer and a Pickering-wax emulsification step, to yield large quantities of photo-responsive active Janus particles. The polymer presents both silane and nitrocatechol groups, binding silica microspheres to a range of metal oxide nanoparticles. The Pickering-wax emulsions protect part of the microspheres' surface, enabling asymmetric functionalization, as required for self-propulsion. The resulting photocatalytic microrobots display a characteristic orientation-dependent 3D active motion upon UV illumination, different to that conventionally described in the literature. By connecting the extensive library of heterogeneous nanoparticle photocatalysts with the nascent field of active matter, this versatile material platform lays the groundwork towards designer microrobots, which can swim by catalysing a broad range of chemical reactions with potential for future applications. 
\end{sciabstract}


\section*{Introduction}

Machines increasingly perform tasks once believed the prerogative of biological systems. Yet, advances in automation are mostly limited to large machines such as cars, drones, and industrial robots. The advent of microrobots, artificial objects that convert energy sources such as light or chemicals into directed motion\cite{Popescu2020}, promises to bring automation down to the micro- and nano-scale. Microrobots can transport matter at the microscale, mix and pump fluids without external agitation \cite{Yuan2021}, thus offering tantalizing opportunities for performing autonomous tasks at small scales in applications ranging from targeted drug delivery \cite{Diez2021}, to environmental remediation \cite{Wang2019}, and even energy conversion \cite{Singh2015}. However, significant technological hurdles must first be overcome if microrobots are to realise their potential for real-world applications. 

Fabrication remains perhaps the most significant challenge to applied microrobotics, as both top-down and bottom-up approaches suffer from significant limitations with respect to scalability and modularity in combining different materials \cite{Wang2020}. Designing novel multi-functional micromachines requires increasingly expensive and specialized equipment, such as for two-photon polymerization \cite{Hu2021}, with implications for the accessibility of scientific research  \cite{MaiaChagas2018}. In contrast to the complex microrobots produced by such techniques, Janus microswimmers are arguably the simplest class of synthetic active matter. These rely on surface patches with different physicochemical properties for propulsion in self-generated gradients \cite{Popescu2020}. In particular, Janus chemical swimmers do not need external actuation, requiring only a chemical fuel source to move. Nevertheless, these simple microrobots often suffer from a very low fabrication throughput due to the methodology used to produce the surface patches \cite{Zhang2019}. 

Janus microswimmers are typically produced by sputter-coating particle monolayers, exploiting line-of-sight vapour-phase deposition and particle self-shadowing  \cite{Wang2020}. Metal coatings are thus selectively deposited as a spherical cap, whose extension can be controlled by tilting the monolayer \cite{Pawar2009,Archer2015}. Despite its widespread popularity, this approach has clear downsides, namely a yield on the order of micrograms and a highly inefficient use of metal precursors. Recently, Archer et al. demonstrated the scalable fabrication of Janus microrobots by functionalizing Pickering-wax emulsions with a two-step nanoparticle seeding and film-growth protocol \cite{Archer2018}. This represents a significant advance in the state-of-the-art of microswimmer fabrication, but the technique is specific to platinum and does not provide close control over the film morphology or composition. The constrained material and synthetic options in the literature, although inconsequential for fundamental studies, hamper the progress of applied microrobotics and, in particular, inhibit the development of propulsion mechanisms based on useful chemical reactions.  

Here, we demonstrate a modular approach to achieve the asymmetric functionalization of microparticles from the toposelective attachment of different nanoparticle thin films, thereby obtaining large (100 mg) quantities of photo-responsive Janus microrobots. Specifically, commercial nanoparticles are asymmetrically attached to SiO\textsubscript{2} microparticles, which are partially embedded in Pickering-wax emulsion droplets, via a poly(acrylamide) modified with silane and nitrocatechol groups \cite{Serrano2016}. The approach is not only scalable but also connects the vast literature on high-surface-area nanocatalysts \cite{Astruc2005,Leeuwen2020} to the fabrication of microswimmers, extending the range of targetable reactions and enabling the facile introduction of new functionalities. 

In particular, by utilizing silane and metal oxide-nitrocatechol chelation chemistry, we demonstrate Janus particles functionalized by TiO\textsubscript{2}, SrTiO\textsubscript{3}, and Fe\textsubscript{2}O\textsubscript{3} nanoparticles.
Furthermore, using a post-modifiable poly(pentafluorophenylacetate) (pPFPAC) backbone presents the opportunity to exploit other metal-coordination chemistries through the introduction of various functional groups. Focusing on commercial TiO\textsubscript{2} P-25 nanocatalysts, we fabricate photoresponsive microswimmers, which not only self-propel under UV-illumination, but also exhibit an interesting orientation-dependent motion that could be exploited for navigation or directed transport \cite{Uspal2019}. 

\section*{Results}

\paragraph*{Polymer-assisted nanoparticle attachment on microspheres} 

The first step in the fabrication of our photocatalytic microrobots is to identify a protocol for the robust attachment of nanoparticle thin films onto the microparticle supports. Electrostatic attachment is frequently used as a means for colloidal heteroaggregation, however, it cannot withstand the harsh cleaning protocols required to remove the solidified wax in Pickering-wax emulsions. Alternative strategies employing covalent bonds are therefore to be sought. In particular, silane groups are frequently used as anchors for SiO\textsubscript{2}, while a raft of coordination chemistries exist for transition metal complexes. However, combining all these features into a single, facile, and robust protocol presents significant challenges. Here, we overcome these obstacles by means of a polymer bridge. This comprises an acrylamide polymer backbone functionalized with both nitrocatechols and silanes to provide multiple covalent linking sites that can stably bind the nanoparticles to the SiO\textsubscript{2} surface. 

We base our polymer bridge on (poly)pentafluoroacetate (pPFPAC), a post-modifiable polymer containing reactive ester linkages, which can be exchanged with amine-containing functional groups by nucleophilic substitution. To decorate our microparticles with various metal-oxide nanoparticles, we functionalize this pPFPAC backbone  with: i) N-boc-hexanediamine, ii) nitrocatechols, and (iii) silane-based groups to obtain poly(acrylamide)-g-(1,6-hexanediamine, 3-aminopropyldimethylethoxysilane, nitrodopamine) \cite{Serrano2016}. The N-boc protecting group on the diamine is removed with TFA, exposing protonated amines on the polymer backbone.  This electrostatic component assists the polymer conformation when binding to negatively charged inorganic surfaces. The silane groups covalently bind the polymer to the SiO\textsubscript{2} microparticle support via siloxane groups, while nitrodopamine facilitates chelation to a range of metals, including titanium and iron oxides \cite{Gulley-Stahl2010,Xu2013}. In this way, the multi-functional polymer acts as a bridge that provides anchoring between the cores of silica microparticles and oxide nanoparticles in a simple heteroaggreation process.

\begin{figure} 
    \centering
    \includegraphics[width=0.75\textwidth]{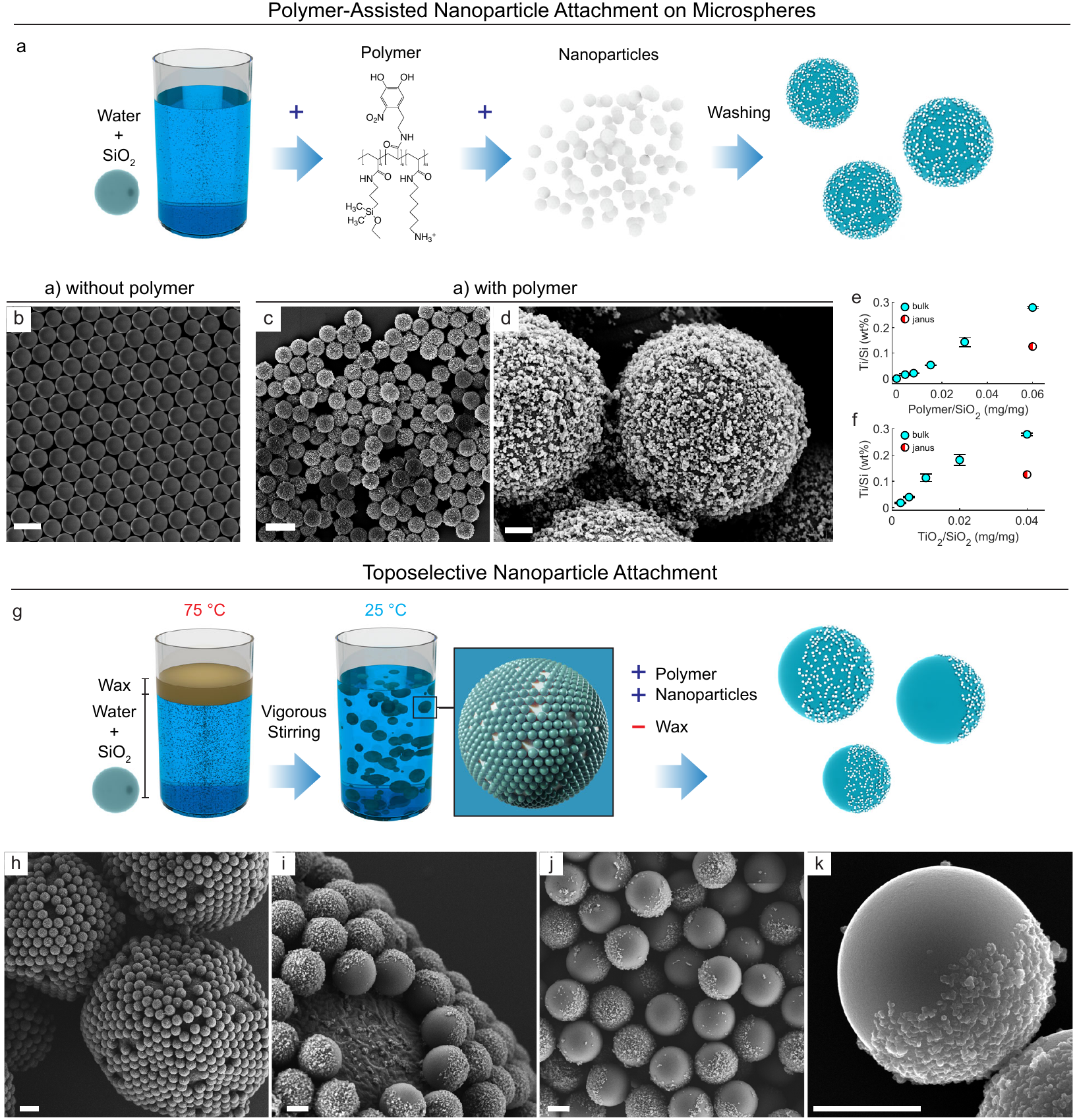}
    \caption{Overview of nanoparticle attachment and combination with Pickering-wax emulsions to achieve Janus particles. a) Protocol to attach nanoparticles to microparticle supports from the sequential mixing of the polymer and nanoparticle suspensions. b) SEM image showing the absence of nanoparticles on SiO\textsubscript{2} without intermediate polymer functionalization step (scale bar 3 $\mu$m). c,d) SEM images showing the successful attachment of TiO\textsubscript{2} P-25 to polymer-functionalized SiO\textsubscript{2} microparticles (scale bars 4 $\mu$m, 0.4 $\mu$m respectively). e) Loading curve of Ti/Si wt\% as a function of polymer added with constant TiO\textsubscript{2} concentration. The amount of added polymer is limited by its solubility in water. f) Loading curve of Ti/Si wt\% as a function of TiO\textsubscript{2} added with constant polymer concentration. g) Pickering-wax emulsion protocol to toposelectively attach pre-synthesized nanoparticles. Wax is used as a temporary mask to screen full coverage of the SiO\textsubscript{2} microparticles. h) SEM image of SiO\textsubscript{2}-wax colloidosomes produced after emulsification (scale bar 5 $\mu$m). i) SEM image showing TiO\textsubscript{2} nanoparticle-functionalized SiO\textsubscript{2} microparticles on SiO\textsubscript{2}-wax colloidosomes (scale bar 1 $\mu$m). j,k) SEM images of Janus particles obtained after removal of the wax. All scale bars are 1 $\mu$m.}
\end{figure}

We optimize the attachment of uniform nanoparticle films focusing on TiO\textsubscript{2} P-25 as our benchmark nanoparticle system, in concurrence with its status within the field of photocatalysis. The process, schematized in Fig. 1a, starts with the activation of the surface of the SiO\textsubscript{2} microparticles by an initial cleaning step, using a hot ammonia and hydrogen peroxide bath to provide available hydroxyl groups to form siloxane bonds with the polymer. The polymer is first dispersed at varying concentrations at $50^\circ$ C overnight. Cleaned particles are then added dropwise to the polymer solutions under magnetic stirring, and left to mix overnight at a final concentration of 0.1w/v\%. After stirring, the SiO\textsubscript{2} is then washed by centrifugation with double-distilled water to remove excess polymer from solution. The SiO\textsubscript{2} particles retain a yellow color from the polymer, due to the presence of nitrodopamine. The functionalized polymer-SiO\textsubscript{2} particles are then redispersed in a PBS 7.0 buffer solution, adapting the protocol of Serrano et al. for flat substrates \cite{Serrano2016}. The particles obtain a pinkish hue, likely arising from conformational changes of nitrodopamine in alkaline environments \cite{Cooper1936}. TiO\textsubscript{2} P-25 of varying concentrations is then added dropwise to a stirred 0.1 w\% solution of the polymer/SiO\textsubscript{2} microparticles and then left mixing overnight. Finally, the P-25/SiO\textsubscript{2} microparticles are washed extensively with alternating sonication and centrifugation steps to remove any excess TiO\textsubscript{2} P-25 not bound to the SiO\textsubscript{2}. 

The produced P-25/SiO\textsubscript{2} microparticles are imaged using SEM. We find that in the absence of the polymer functionalization step, no TiO\textsubscript{2} P-25 nanoparticles are bound to the SiO\textsubscript{2} microparticles after washing (Fig. 1b), even at the highest TiO\textsubscript{2} concentrations investigated. We also verify this with ICP-OES elemental analysis, and find negligible quantities of Ti in the sample (Fig. 1e), highlighting the effectiveness of the cleaning protocol and the requirement for the polymer to bind the TiO\textsubscript{2} P-25 nanoparticles (Figs. 1b-e). The TiO\textsubscript{2} loading can be effectively controlled by varying the amount of polymer and TiO\textsubscript{2} P-25 added, and is retained after washing (Figs. 1e,f). We observe a linear growth in the Ti/Si ratio with increasing polymer concentration (Fig. 1e), which is limited by the solubility of the polymer in water. The Ti/Si ratio would naturally saturate with increasing TiO\textsubscript{2} P-25 (Fig. 1f), but, at the highest TiO\textsubscript{2} concentrations we explore, aggregation of the nanoparticles starts to occur. 

\paragraph*{Toposelective nanoparticle attachment} 
Having established the success of the bulk surface modification, we combine it with a Pickering wax emulsion approach to produce asymmetrically functionalized SiO\textsubscript{2} microparticles, en route to realizing photocatalytic microrobots \cite{Hong2006}. This strategy consists in decorating the surface of molten wax droplets in an aqueous medium with SiO\textsubscript{2} microparticles. The particles are irreversbily adsorbed at the water-wax interface and are immobilized when solidifying the wax upon cooling. The surface of the particles immersed in the wax is then protected from surface modifications that are carried out in the aqueous medium. In particular, we prepare our Pickering emulsions by adapting the methodology described by Perro et al. \cite{Perro2009}. Cleaned particles are dispersed in didodecyldimethylammonium bromide (DDAB) solutions with concentrations corresponding to an approximate surfactant monolayer coverage on all particles \cite{Kalai2019}. Wax is added to the suspension, which is heated to $75^\circ$ C, and then subjected to a two-step vigorous stirring protocol \cite{Lebdioua2018} (see experimental section for more details). The hot emulsion is then rapidly cooled in an ice-bath to obtain solidified SiO\textsubscript{2}-wax colloidosomes (Figs. S1a-d). The colloidosomes are then washed consecutively by gravitational sedimentation with distilled water, a 0.1 M NaCl solution to remove the DDAB cationic surfactant, then water once more to remove the salt. At this stage, we add the multifunctional polymer to coat the exposed surface of the SiO\textsubscript{2} particles. The colloidosomes are dispersed in the polymer solution, gently agitated overnight with an orbital mixer, then washed once more with distilled water to remove excess polymer, redispersed in a pH 7.0 PBS solution, and gently mixed with the nanoparticles overnight. Finally, the colloidosomes are collected and dried, before mixing and filtering with chloroform to remove the wax (Fig. 1g).

For straightforward functionalization, the colloidosomes should be denser than water (Fig. S3). Colloidosomes with a mean diameter of 33.4 $\mu$m are produced using 2.12 $\mu$m particles (Supplementary Fig. 2). A particle concentration of 5w/v\% in water is emulsified with wax in a 1:10 wax:water volumetric ratio. The process can be readily tuned to produce colloidosomes with SiO\textsubscript{2} particles of varying sizes (Figs. S1a-d). Microparticle size provides a convenient handle on controlling physical properties of the final microrobots such as the rotational diffusion coefficient, which could be exploited for e.g. enhanced mixing \cite{Lin2011} or navigation \cite{Fernandez-Rodriguez2020}. However, unlike some previous reports, we were not able to tune particle penetration into the wax with surfactant concentration - instead we found that the surfactant concentration only determined whether or not it was possible to obtain Pickering emulsions \cite{Kalai2019}.  

\begin{figure}
    \centering
    \includegraphics[width=0.5\textwidth]{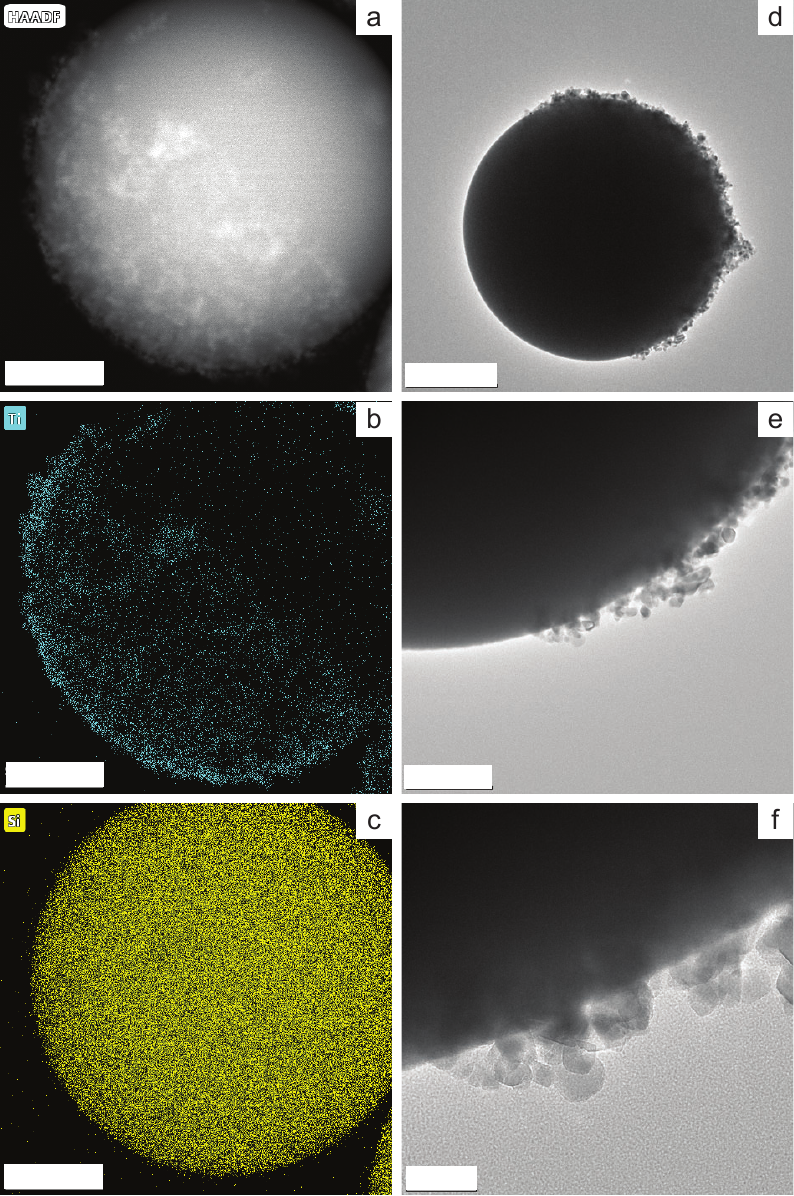}
    \caption{HR-TEM and elemental mapping of asymmetric TiO\textsubscript{2} P-25 nanoparticle thin films on SiO\textsubscript{2} microparticles. a-c) HAADF-STEM and EDS mapping of Janus particles: the asymmetric functionalization with TiO\textsubscript{2} is clearly visible (scale bars 500 nm). d-f) HR-TEM of the nanoparticle films indicate direct attachment of porous aggregate structures (scale bars 750 nm, 200 nm, 50 nm respectively).}
\end{figure}

Our process gives an approximate 50-50 TiO\textsubscript{2}/SiO\textsubscript{2} surface coating (Figs. 1i-k). Ti loading is also confirmed by ICP-OES, which gives an approximate 50\% coverage for the Janus microrobots, as observed with SEM (Figs. 1e,f). The value contrasts with the expected nanoparticle coverage based on the penetration of the SiO\textsubscript{2} microparticles into the wax (approximately 0.36R from direct measurement of the three-phase contact angle) and on previous findings using Pickering-wax emulsions \cite{Archer2018,Hong2006}. We hypothesize that the closely packed monolayers of SiO\textsubscript{2} form an effective barrier to the transport of TiO\textsubscript{2} P-25 nanoparticles onto the whole surface of the silica particles that is not protected by the wax, thereby preventing a higher surface coverage. The rotated particles in Fig. 1i, likely a result of subsequent filtration steps after nanoparticle attachment, also suggest this shadowing effect. The Janus morphology is retained after the harsh cleaning protocols necessary to remove excess nanoparticles and wax, and the redispersal of the dried Janus microrobots in water (Figs. 1j,k), indicating the durability of the polymer linkage. 

We confirm the Janus distribution of TiO\textsubscript{2} by elemental mapping, namely HAADF-EDS (Figs. 2a-c). Using TEM, we are also able to visualise the morphology of the thin nanoparticle films. The thin films are networks of attached nanoparticle clusters formed from multiple TiO\textsubscript{2} primary particles rather than individual nanoparticle structures. This is in agreement with the expected morphology of commercial TiO\textsubscript{2} P-25 \cite{Ohno2001}. The formation of such porous nanoparticle structures is favourable due to their enhanced surface area compared to dense films \cite{Gao2015}, thereby increasing catalytic activity and thus swimming speeds \cite{Choudhury2015}.

Utilizing nitrocatechol chelation chemistry extends the applicability of our polymer-based nanoparticle attachment to a range of metal oxides, and could therefore be exploited to obtain composite thin films of functional nanoparticles on microparticle supports. To this end, and to demonstrate the generality of our method, we also attach different phases of TiO\textsubscript{2} (amorphous, anatase, and rutile), Fe\textsubscript{2}O\textsubscript{3}, and SrTiO\textsubscript{3} using the same protocols developed for TiO\textsubscript{2} P-25 (Fig. 3). Fe\textsubscript{2}O\textsubscript{3} imparts both photocatalytic and magnetic properties, and could be combined with TiO\textsubscript{2} P-25 for enhanced speeds \cite{Maric2020} and controlled steering \cite{Sridhar2020a}. SrTiO\textsubscript{3} is a perovskite photocatalyst widely studied for its water-splitting properties \cite{Goto2018}, and therefore is promising as the basis for a fuel-free microswimmer. Toposelective nanoparticle attachment is thus a promising modular route to obtaining large quantities of microrobots with a range of functionalities that can be tuned by selection of the starting materials.  

\begin{figure}
    \centering
    \includegraphics[width=\textwidth]{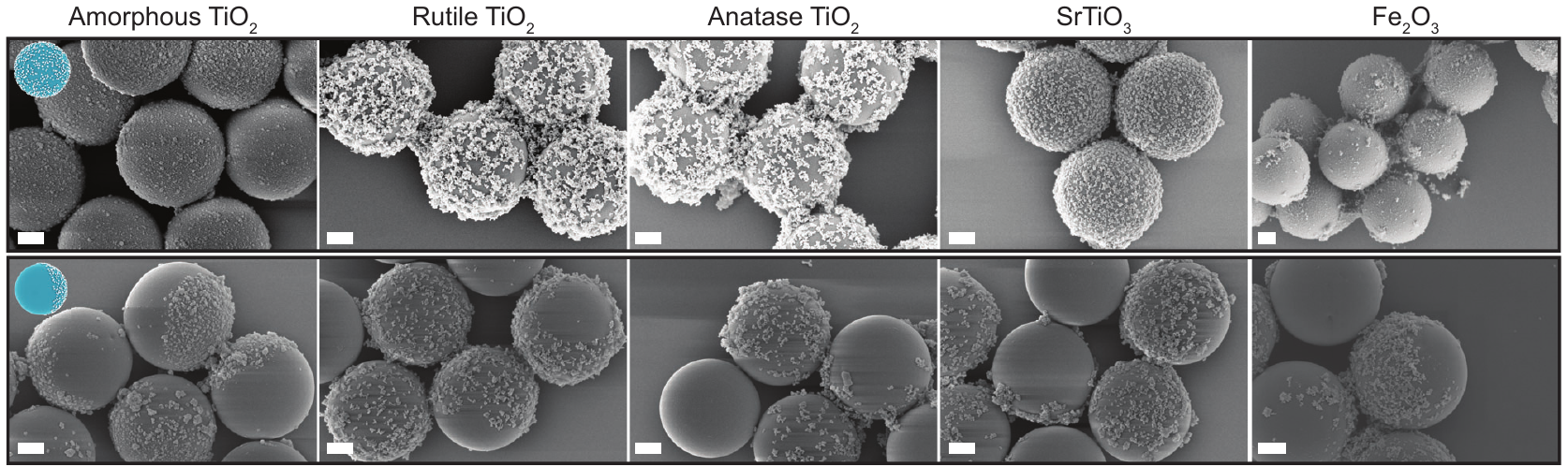}
    \caption{Symmetric and asymmetric functionalization of SiO\textsubscript{2} microparticles with various commercial, pre-synthesized nanoparticles, highlighting the versatility of the proposed approach (scale bars 500 nm).}
\end{figure}

\paragraph*{Microrobots from micro- and nano-particles swimming in 3D} 

To investigate the autonomous, photocatalytic motion of our TiO\textsubscript{2} P-25 microrobots, we perform single-particle tracking studies with bright-field microscopy, using in-house particle-tracking scripts (Fig. S8), under different illumination conditions and H\textsubscript{2}O\textsubscript{2} concentrations. TiO\textsubscript{2} is known to degrade H\textsubscript{2}O\textsubscript{2} under UV light, and the Janus distribution of TiO\textsubscript{2} P-25 nanoparticles on the SiO\textsubscript{2} core's surface thus leads to the formation of asymmetric gradients around the microswimmers. These in turn develop flow fields which result in self-diffusiophoresis of the particles \cite{Anderson1989} (Figs. 4a,b). We first confirmed the photo-responsive behavior of the microrobots by alternating off-on UV illumination cycles and found that the UV illumination is a necessary pre-condition for motion (Figs. 4c,d), with no evidence of memory or photo-charging effects \cite{Sridhar2020}. We then evaluated the instantaneous velocities of the microrobots under different illumination strengths and wavelengths (Fig. S8).

The trajectories of microswimmers are typically described with a 2D active Brownian motion model \cite{Dietrich2017}, where a constant propulsion velocity is randomized in 2D by rotational diffusion. More recently, there has also been a focus on fabricating and studying microswimmers with unbounded 3D active motion \cite{Yasa2018}. However, our microrobots demonstrate more complex behavior, which does not follow the 2D and 3D active Brownian motion equations \cite{Loewen2020}. Specifically, their motion is mostly confined to the 2D plane, with interdispersed short and rapid periods of out-of-plane motion (Figs. 4e,h). The predominantly 2D motion may be explained by hydrodynamic interactions with the glass substrate, which favor in-plane motion \cite{Uspal2015a}. Competing effects from out-of-plane rotational diffusion and angular velocity arising from a non-uniform nanoparticle coating and  wall-induced flows \cite{Ruhle2018} could cause the observed random out-of-plane ballistic segments. To characterise this 3D motion, we first confirm that in the absence of chemical fuel the particles are able to rotate in 3D (Video S2) and measure their rotational diffusion in solvents of varying viscosity in line with theory (Table S1) \cite{Anthony2006}. We then measure their 3D active trajectories via a simple approach making use of the changing diffraction patterns of the particles as they swim in and out of the focal plane (see Supplementary Text for details). 

\begin{figure}
    \centering
    \includegraphics[width=\textwidth]{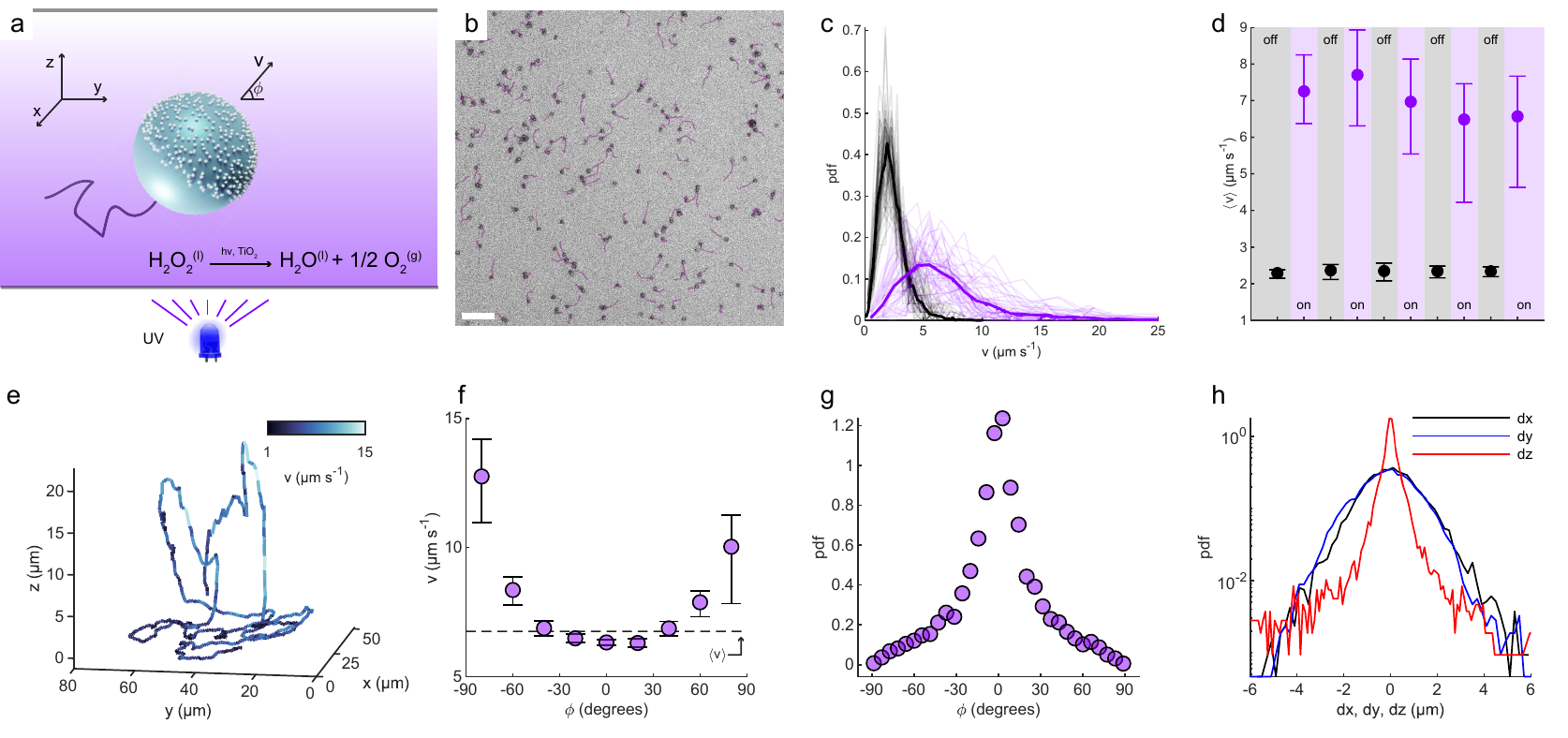}
    \caption{Overview of particle motion under UV illumination in fuel-rich media. a) Schematic of particle motion. Under the decomposition of hydrogen peroxide, the microrobots swim with the functionalized cap forwards. b) Wide-field image with superimposed trajectories (shown for 3 s) under UV illumination (scale bar 20 $\mu$m). c) Distribution of instantaneous velocities during off (black) and on (violet) cycles. d) Mean instantaneous velocities of particles during alternating off-on cycles of UV illumination. e) Example of a microrobot trajectory in 3D, with the magnitude of the velocity vector color-coded to illustrate the occurrence of an orientation-dependent velocity. f) Plot of particle velocity vs. out-of-plane orientation angle. Particles swim faster out of plane, and the fastest motion is observed when the particle swims towards the glass substrate from above ($\phi = -90^\circ$). g) Distribution of out-of-plane  orientation angles ($\phi$) across all particle trajectories. h) Distribution of displacements for all particles in x, y, and z. The distributions in x and y (in-plane) are Gaussian and similar, while the distribution of z displacements shows pronounced tails. Error bars correspond to 95\% confidence intervals, which were obtained with bootstrapping.}
\end{figure}

From the analysis of individual 3D trajectories, like the one reported in Fig. 4e, we measure median instantaneous velocities ranging from 2-13 $\mu$ms\textsuperscript{-1} on a per particle basis, and note that the particles swim cap first (Fig. 4a, Video S1). Under the conditions found to maximize swimming speeds (3v\% H\textsubscript{2}O\textsubscript{2}, 340 nm, 9.4 mWmm\textsuperscript{-2}), we find a median instantaneous velocity of approximately 6.5 $\mu$ms\textsuperscript{-1} on a per particle basis (Fig. S8). Lowering the fuel concentration leads to lower swimming speeds \cite{Howse2007}, and we also note a difference in swimming speed depending on the wavelength of illumination used. 

By plotting the instantaneous velocities as a function of particle orientation, we observe a significant asymmetry in the particle motion (Figs. 4f-h). In particular, the microrobots exhibit orientation-dependent velocities, displaying faster motion as their orientation is increasingly directed out of plane. The effect persists over multiple frames, indicating that it is not a tracking artifact. Moreover, contrary to expectations where shadowing dictates particle motion \cite{Singh2018}, the observed faster segments are not uni-directional. Furthermore, the previously discussed on-off photo-responsiveness of the particles excludes a memory or charging effect \cite{Sridhar2020}, which might have explained the bi-directional fast swimming segments (Figs. 4c,d). Based on the fact that the transmission of 340 nm wavelength light through 2 $\mu$m fused SiO\textsubscript{2} is on the order of 90\% (even after accounting for Fresnel losses), we hypothesise that the SiO\textsubscript{2} core does not block the UV light. This implies that even when the catalytic cap is completely "shadowed", upwards swimming is observed, and we attribute the fastest motion towards the substrate (cap down) to the direct illumination of the catalytic cap and the concurrent effect of gravity. The slower in-plane motion could be caused by the orientation of the propulsion direction, and the near-wall hydrodynamic interactions which increase drag forces \cite{Goldman1967}. Although statistics for 2D active Brownian trajectories with orientation-dependent velocities have been reported \cite{Sprenger2020}, their extension to the the 3D case reported here presents potential for future theoretical developments.

\section*{Discussion}

We have demonstrated a modular approach to fabricate large (100 mg) quantities of photo-responsive microrobots from the asymmetric attachment of commercial nanoparticles. Our method for toposelective nanoparticle attachment provides a versatile material platform which can be extended to various functional nanoparticles. We envision that it will offer an off-the-shelf modular approach to obtain large quantities of Janus particles with mixed composite films for a range of applications. The described protocol does not require specialised equipment beyond that found in typical synthesis laboratories, lending itself to wide-spread application. 

Moreover, the motion of our photo-responsive microrobots highlights several avenues for further research. Analysis of the observed 3D swimming behaviour requires a new theoretical framework not currently found in the literature. We also note that the TiO\textsubscript{2} P-25 microrobots self-propel more slowly at higher illumination wavelengths, which we attribute to TiO\textsubscript{2}’s large energy band gap. This demonstrates the opportunities to exploit the wealth of catalysis literature \cite{Etacheri2015} to inform the design of microrobots with desirable attributes (e.g. with faster swimming speeds or visible light activation) by appropriate selection of nanocatalytic materials.

Finally, the ballistic out-of-plane motion of our photocatalytic microrobots suggests a photoreactor design where the competition between gravity and activity is exploited. Realizing motion control in 3D, for example by dynamic light patterning \cite{Arrieta2019}, could induce complex flows, which in turn can enhance overall reactor efficiencies in traditionally difficult-to-mix settings. Such control would not only be favorable for mixing in microfluidic channels \cite{Ward2015}, which have similar dimensions to the observed Z displacements of the microswimmers, but also in flat-panel reactors where mass transfer is a key limiting factor \cite{Takanabe2017}. Increased reaction rates arising from microswimmer motion have been previously demonstrated \cite{Orozco2013}, but applicability has been limited by materials and scalability. By targeting societally relevant reactions, such as water splitting, we hypothesise that scalable microrobots could be exploited in a novel photoreactor concept where the particles possess a dual catalyst-stirrer functionality. Therefore, incorporating aspects of chemical reaction engineering and soft matter physics could help overcome the four-phase mass-transfer limitations inherent to current photocatalytic systems.

\section*{Materials and Methods}
\paragraph*{Nitrodopamine Synthesis}
Nitrodopamine was synthesized following well-established  protocols \cite{Napolitano1992}. Briefly, dopamine(5g, 32.6 mmol) and sodium nitrite (6.3g, 91.3 mmol) were dissolved in 150 mL water and cooled to $0^\circ$ C under stirring in an ice bath. 25mL sulfuric acid (20v/v\%) was added dropwise and left stirring at r.T. overnight. The product mixture was cooled once more to $0^\circ$ C, filtered, and washed with copious amounts of double-distilled water at $0^\circ$ C and then ethanol at $0^\circ$ C. The resulting nitrodopamine hydrogen sulfate was then dried under high vacuum overnight.  

\paragraph*{poly(acrylamide)-g-(1,6-hexanediamine, 3-aminopropyl-dimethyloxysilane, nitrodopamine) synthesis}
Synthesis of the pentafluorophenyl acrylate monomer and its polymerization were performed following previously published protocols \cite{Serrano2016}. Briefly, pentafluorophenol (87.21 g, 0.47 mol) was dissolved in 150 mL of dichloromethane (DCM) at $0^\circ$ C and 2,6-dimethylpyridine (60.55 mL, 0.52 mol) was added slowly through a dropping funnel, which was afterwards rinsed with another 150 mL of DCM. This second portion was added to the reaction mixture. Acryloyl chloride (42.14 mL, 0.52 mol) was then added dropwise to the reactor, still under cooling, and left to react overnight under N\textsubscript{2} atmosphere at room temperature. The resulting 2,6-dimethylpyridine hydrochloride salt was removed by filtration and the residual solution was washed three times with 100 mL of water, dried with magnesium sulfate and the solvent evaporated under reduced pressure. The product monomer was purified by distillation (in two portions) under reduced pressure (10 mbar) to give a colorless liquid (97.09 g, 78\%). The monomer pentafluorophenyl acrylate (14.31 g, 60.13 mmol), the initiator AIBN (23.83 mg, 0.15 mmol) and the chain-transfer agent 2-(dodecylthiocarbonothioylthio)-2-methylpropionic acid (158.45 mg, 0.43 mmol) were dissolved in 15 mL of toluene inside a Schlenk tube. The solution was degassed via three freeze-pump-thaw cycles and left to react under a nitrogen atmosphere at $80^\circ$C in an oil bath for 18h. After the RAFT polymerization was completed, the mixture was left to cool to room temperature and the resulting polymer (pPFPAC) isolated by precipitation in methanol and dried under vacuum for 48h (12.90 g, 90\%).

Likewise, the post-modification steps were carried out as outlined in the work by Serrano et al., with the exception that the (poly)pentafluoroacetate (pPFPAC) backbone was not first PEGylated, and instead is only post-modified with the binding side groups (N-boc-hexanediamine, 3-aminopropyldimethylethoxysilane, nitrodopamine). Briefly, N-boc-hexanediamine (227 mg, 1.05 mmol) was dissolved in 6.4 mL dimethylformamide (DMF) with an excess of triethylamine (318 mg, 3.15 mmol). The mixture was added drop-wise to pPFPAC (500 mg, 2.1 mmol) dissolved in 5.07 mL DMF and left stirring overnight at $50^\circ$ C. A new solution containing 84.5 mg (0.525 mmol) of 3-aminopropyldimethylethoxysilane and triethylamine (160 mg, 1.58 mmol) was dissolved in 7.4 mL of DMF and added drop-wise to the previous solution, still at $50^\circ$ C and under stirring overnight. An excess of nitrodopamine was dissolved separately (154.4 mg, 0.525 mmol) in 7.4 mL of DMF with 160 mg of triethylamine (1.58 mmol). The latter solution was added slowly to the polymer solution and left stirring overnight at $50^\circ$ C. DMF was evaporated under reduced pressure, the mixture re-dissolved in DCM (40 mL, 4 equivalents) and trifluoroacetic acid (TFA, 10 mL, 1 equivalent) and left to react under stirring overnight. The resulting mixture was again evaporated under reduced pressure and re-dissolved in twice-distilled water (80 mL). This solution was purified by dialysis against water for two days using a membrane with a MWCO of 3,500 Da and subsequently freeze-dried to obtain the yellow-brown poly(acrylamide)-g-(1,6-hexanediamine, 3-aminopropyl-dimethyloxysilane, nitrodopamine). 

\paragraph*{Polymer-assisted nanoparticle attachment onto microspheres}
Polymer solutions were prepared by dispersing the dry poly(acrylamide)-g-(1,6-hexanediamine, 3-aminopropyl-dimethyloxysilane, nitrodopamine) in $50^\circ$ C water overnight under magnetic stirring. The maximum polymer concentration is limited by the solubility of the polymer (approximately 60 mg/L). 1 w/v\% SiO\textsubscript{2} microparticles suspensions were first added to a bubbling $70^\circ$ C H\textsubscript{2}O\textsubscript{2}/NH\textsubscript{4}OH solution (1:1:1 volumetric ratio) under magnetic stirring for 10 minutes to activate the SiO\textsubscript{2} surface with reactive hydroxyl groups. The cleaned particles were then added dropwise under magnetic stirring to the prepared polymer solutions and left stirring overnight (final SiO\textsubscript{2} concentration 0.1 w/v\%). The polymer-SiO\textsubscript{2} particles were then washed by centrifugation to remove excess polymer and redispersed in phosphate-buffered saline (PBS pH 7.0). Nanoparticle suspensions of varying concentrations ((PBS pH 7.0 media) were then added dropwise to the polymer-SiO\textsubscript{2} suspensions under magnetic stirring and left mixing overnight (final SiO\textsubscript{2} concentration 0.1 w/v\%). Finally, the nanoparticle functionalized SiO\textsubscript{2} microparticles were washed extensively with alternating sonication and centrifugation steps to remove any excess nanoparticles not bound to the SiO\textsubscript{2} microparticles. 

\paragraph*{Microrobot Fabrication}
Wax-SiO\textsubscript{2} Pickering Emulsions were prepared by adapting the methodology used by Perro et al \cite{Perro2009}. Suspensions containing 5w/v\% SiO\textsubscript{2} particles, 10.8 mg/L didodecyldimethylammonium bromide (DDAB) \cite{Lebdioua2018}, and a 1:10 molten wax: water volumetric ratio were heated to $75^\circ$ C, then stirred for 15 minutes at 3000 RPM before vigorous mixing at 13500 RPM for 160s using an IKA T-25 Digital Ultraturrax \cite{Chu2020}.  After the emulsification step, the Pickering emulsion was immediately placed in an ice bath to rapidly solidify the colloidosomes. The emulsion was then washed in a 0.1M NaCl solution to remove surfactants, before further washing in deionised water. The SiO\textsubscript{2}-Wax colloidosomes were dispersed overnight by gentle agitation in an aqueous solution of a post-modified (poly)pentafluoroacetate (pPFPAC) polymer. The pPFPAC-colloidosomes were then washed thoroughly in deionized water before redispersion in a phosphate-buffered saline (PBS) pH 7.0 suspension containing the functional metal-oxide nanoparticles. After gentle mixing overnight, the nanoparticle functionalized colloidosomes were collected by filtration and the wax is finally removed with chloroform. 

\paragraph*{Light-controlled motion experiments}
Stock H\textsubscript{2}O\textsubscript{2} (30 v/v\%, manufacturer) was added to dilute particle suspensions of the microrobots to obtain 300 uL of the desired H2O2 concentrations. 280 uL thereof was then pipetted into a flow-through cell (cell 137-QS; Hellma Analytics) with a light path length of 1 mm. Particles were imaged on an inverted microscope under Köhler illumination using a 40x objective (CFI S Plan Fluor ELWD 40XC) with adjustable collar (set to 1 mm), and videos were taken at 10 fps on a Hamamatsu C14440-20UP digital camera. UV illumination of the particles (340/380 nm) was achieved using a Lumencor SPECTRA X light engine as the excitation source through the objective.  

\paragraph*{Particle Tracking}
Videos were first pre-processed with Fiji (ImageJ) for conversion to 8-bit and cropped to obtain one particle per field of view. Particles centres were tracked using a custom script combining the MATLAB implementation of the Crocker and Grier IDL particle tracking method \cite{Crocker1996}, and the radial symmetry approach outlined by Parthasarathy \cite{Parthasarathy2012}. The first invariant moment (inertia) of masks around the particle centres was determined using a MATLAB implementation of Hu’s 7 invariant moments formulation \cite{Hu1962}. Inertia look-up-tables (LUTs) were first obtained for stationary and diffusive particles. The evolution of inertia with Z was then inverted, before fitting of a cubic polynomial using an inbuilt MATLAB non-linear regression function (nlinfit). From this cubic functional form, the Z values of active particles could be determined with prediction intervals from inertia of their masks (see Supplementary Text for more details).

\bibliography{Paper1_Janus_TNA.bib}

\bibliographystyle{Science}

\section*{Acknowledgments}
The authors thank Dr. T Gmür and Dr. K Zhang for their assistance with polymer synthesis and characterisation. They are also grateful to Dr. S.S Lee for his assistance with the UV microscopy set up used for the particle tracking experiments, to Dr. P Zeng for the images taken with SEM and TEM, to V Niggel for the discussions on image analysis and lab safety, and to Dr. B Hattendorf for the ICP-OES measurements. M.R Bailey would additionally like to thank Dr. G.A Filonenko for his various insights pertaining to the synthesis. We acknowledge the Scientific Center for Optical and Electron Microscopy (ScopeM) of ETH Zurich for access to their instrumentation and expertise. Fabio Grillo acknowledges financial support from the Swiss National Science Foundation, grant number 190735.

\paragraph*{Author Contribution Statement} 
Author contributions are defined based on the CRediT (Contributor Roles Taxonomy). Conceptualization: M.R.B., F.G., L.I. Formal Analysis: M.R.B., F.G. Funding acquisition: F.G., L.I. Investigation: M.R.B. Methodology: M.R.B., F.G., L.I., N.S. Software: M.R.B., F.G. Supervision: F.G., L.I. Validation: M.R.B., F.G. Visualization: M.R.B., F.G., L.I. Writing - original draft: M.R.B., F.G., L.I. Writing - review and editing: M.R.B., F.G., L.I., N.S.   

\clearpage


\newpage
\section*{Supplementary materials}

\begin{itemize}
    \setlength\itemsep{-1em}
    \item[] Text S1: Preparation of SiO\textsubscript{2}-wax colloidosomes
    \item[] Text S2: Polymer characterisation
    \item[] Text S3: 3D tracking with invariant moments 
    \item[] Text S4: Polymer stability
    \item[] Text S5: 3D rotation of microrobots
    \item[] Text S6: Motion under different experimental conditions
    \item[] Fig. S1: Colloidosomes from different particle sizes
    \item[] Fig. S2: Size distribution of colloidosomes
    \item[] Fig. S3: Colloidosome density
    \item[] Fig. S4: \textsuperscript{1}HNMR spectrum
    \item[] Fig. S5: ATR-IR spectra
    \item[] Fig. S6: LUT of Inertia(Z)
    \item[] Fig. S7: Microrobot stability under highly oxidizing conditions
    \item[] Fig. S8: Box-plots of instantaneous velocity under different conditions
    \item[] Table S1: Conditions to obtain colloidosomes
    \item[] Table S2: 3D rotational diffusion of microrobots
\end{itemize}
\newpage
\newpage
\section*{Supplementary Text}
\paragraph*{Text S1: Preparation of SiO\textsubscript{2}-wax colloidosomes}

SiO\textsubscript{2}-wax colloidosomes were prepared by adapting the protocol proposed by Perro et al. as described in the main manuscript \cite{Perro2009}.  In contrast to the methodology presented in \cite{Hong2006}, here the particles are dispersed in the water phase with surfactants, before vigorous stirring by means of an Ultra-Turrax, to obtain the Pickering wax-in-water emulsion. We find that a 5 w/v\% SiO\textsubscript{2} in water concentration provides the best results when preparing the emulsions (5 mL). For the 2.12 $\mu$m SiO\textsubscript{2} particles that are the focus of this work, a volumetric wax:water ratio of 1:10 was used.  Surfactant concentrations were selected such that an approximate monolayer coverage of the SiO\textsubscript{2} particles is achieved \cite{Kalai2019}. The flexibility of the approach is demonstrated by producing emulsions with various particle sizes (Table S1, Fig. S1), suggesting an adaptable approach to obtain microrobots with different physical properties.

We note here the importance of the overall colloidosome density. To ensure effective dispersion of the colloidosomes during functionalization, their density should be greater than that of water. However, paraffin wax has a density of 900 kgm\textsuperscript{-3}, and so colloidosome size can affect dispersibility. As a rough estimation, we find that for 2 $\mu$m particles that fully cover the wax droplet's surface, a maximum colloidosome size of approximately 100 $\mu$m ensures that the colloidosomes are denser than water (Fig. S3). The various washing procedures, which depend on sedimentation, ensure that larger colloidosomes are removed from the product before further functionalization steps. 

\paragraph*{Text S2: Polymer characterisation}
The D\textsubscript{2}O \textsuperscript{1}HNMR spectrum of the post-modified poly(acrylamide)-g-(1,6-hexanediamine, 3-aminopropyl-dimethyloxysilane, nitrodopamine) (Fig. S4) is largely in agreement with literature \cite{Serrano2016}. Chemical shifts of 0.9 to -0.1 and 1.7 to 1.2 ppm, attributed to the presence of 3-aminopropyl-dimethyloxysilane and 1,6-hexanediamine respectively, are observed. Likewise, the peak positions at 3.5 to 2.8 ppm correspond to the polymer backbone. However, we are not able to identify the nitro-dopamine group from the NMR spectrum, although we note that in the work by Serrano et al. the reported signal is also very weak.

To further investigate the functional groups present on the post-modified polymer backbone, ATR-IR measurements were also taken (Fig. S5). The unmodified pPFPAC backbone displays two strong peaks at 990 cm\textsuperscript{-1} and 1520 cm\textsuperscript{-1}, arising from the pentafluorophenyl groups (C-F stretch and C-C aromatic stretch, respectively \cite{Serrano2016}). We note that after post-modification, the C-F stretch disappears, while a new set of peaks at 1500-1550 cm\textsuperscript{-1} emerge. Looking at the spectrum for the synthesized nitrodopamine, we identify the same structure and attribute these multiple peaks to the combined contribution of the aromatic C-C in-ring stretch and NO stretch of nitrodopamine. We also identify; i) peaks which we attribute to 1,6-hexanediamine (3000-2850 cm\textsuperscript{-1}, C-H stretch), ii) multiple peaks in the fingerprint regime characteristic of silanes, and iii) a strong merged peak around 1625 cm\textsuperscript{-1} which is likely a combination of the signal from the C=O stretch and in-plane N-H bend of the secondary amides formed when coupling the functional groups to the polymer. 

\paragraph*{Text S3: 3D tracking with invariant moments}

To characterise the 3D swimming behaviour of the microrobots, we developed a simple image-invariant-based approach to determine the displacements in Z. We exploit the changing diffraction pattern of particles imaged under bright-field Köhler illumination as they move in and out of the focal plane. We quantify this change in the image properties by taking the inertia (first moment) of image masks centred at the particle. By studying trajectories in the Z region where the inertia changes monotonically, we are therefore able to track motion in 3D.

We first create a look-up-table (LUT) of inertia values by obtaining Z-stacks of stationary particles with 0.2 $\mu$m increments (Fig. S6). A circular mask around the particle center is taken for each Z-stack image, from which the first invariant image moment (inertia) is calculated \cite{Hu1962}, \cite{MATLAB_Hu_2021}. By performing this operation for each Z slice, a series of inertia(Z) calibration curves are obtained. These curves are then averaged, and the monotonic domain is inverted to obtain Z(inertia), to which a cubic polynomial is fitted using an in-built MATLAB function (nlinfit). 

This cubic polynomial is then used to determine the Z-position of active swimming trajectories. Briefly, we set the focal plane above the particle centres so that as the particles swim upwards, they move into focus. This is done to maximise the monotonically increasing regime of the particle diffraction ring's image inertia. We then image this plane under UV illumination with H\textsubscript{2}O\textsubscript{2} fuel as for normal 2D particle tracking experiments. For each frame, the image moment of the particle is computed as previously described for the Z-stacks. From the previously fitted cubic polynomial, a prediction for the microswimmer out-of-plane position is thus obtained from Z(inertia) with an average prediction error of approximately +/- 0.5 $\mu$m. The described methodology carries a distinct advantage over more conventional confocal approaches, in that a much higher FPS is achievable since only one Z plane is imaged with wide-field microscopy \cite{Kovari2019}. Such methods also require computationally expensive point spread functions (PSFs) for accurate center finding \cite{Speidel2003}. They are therefore well suited for tracking fluorescent passive particles in suspensions, but less well adapted for tracking photo-responsive asymmetric active colloids in bright-field.

To capture the diffraction rings of the particles consistently, accurate center finding is important for the image masks from which the first invariant moment is calculated.To enhance particle localization and efficiency, the MATLAB implementation of the Crocker and Grier algorithm \cite{Crocker1996} is used as a first approximation to identify the particles. Masks are then created around these particles, to which the Parthasarathy radial symmetry approach \cite{Parthasarathy2012} is applied. By combining the two approaches, we find that center finding is improved with respect to the base cases, and works well when the particles are both in and out of focus.  

\paragraph*{Text S4: Polymer stability}
We investigate the stability of nanoparticle attachment via the polyacrylamide by imaging the microrobots after a simulated particle tracking experiment. The microswimmers are illuminated with high power UV (365 nm, 27 mWmm\textsuperscript{-2}) in a H\textsubscript{2}O\textsubscript{2} rich (5v/v\%) environment for 30 minutes. The conditions used exceed the  experimental conditions used for the single particle tracking experiments, and provide a highly oxidative environment which would degrade many organics. We then collect the particles by centrifugation and image them with SEM. We find no change to the morphology of the TiO\textsubscript{2} nanoparticle thin films (Fig. S7), indicating a stable attachment. 

\paragraph*{Text S5: 3D rotation of microrobots}

By illuminating the particles with a mixture of bright-field and 440 nm wavelength light, we clearly observe 3D rotation of the passive microrobots (see Video S2). This contrasts with descriptions of purely 2D microswimmers possessing an orientation vector diffusing on the unit circle. The observation also indicates that bottom-heaviness likely plays a less significant role for our microrobots than in previous reports \cite{Singh2018}. We expect that our porous TiO\textsubscript{2} nanoparticle films are less heavy than the 55 nm dense TiO\textsubscript{2} films obtained by e-beam evaporation, resulting in less constrained rotation \cite{Uspal2019}.

We use the "MOON" approach proposed by Anthony et al \cite{Anthony2006} to describe this rotation. The centres of the particles, and the centroids of the nanoparticle thin films are detected using in-house python scripts. The vector connecting these structural points defines a diffusing point on a sphere using a spherical coordinate system. The angle of variation between these points can thus be obtained, from which the mean squared angular displacement (MSAD) can be determined. We fit Perrin's equation for small angular displacements at short timescales $ <\omega\textsuperscript{2}> = 4D\textsubscript{R}\Delta t $ to the MSAD to obtain D\textsubscript{R}. As in \cite{Anthony2006}, it is necessary to include an intercept term at $\Delta t = 0$ to obtain a fit of the slope (D\textsubscript{R}), which approximates the uncertainty in $\omega$. D\textsubscript{R} values thus obtained decay with viscosity as expected from theory, indicating a physical origin for the values (Table S2). We note occasional instabilities when finding the cap centroid which can lead to outlier values, however we stress that the video clearly demonstrates 3D rotation of the particles.

\paragraph*{Text S6: Motion under different experimental conditions}

The instantaneous velocities of the P-25 microrobots were evaluated under different experimental conditions (Fig. S8). The concentration of fuel (H\textsubscript{2}O\textsubscript{2}), illumination intensity (mW mm\textsuperscript{-2}), and the wavelength of illumination were varied for dilute suspensions of the microswimmers. We find that decreasing the 340 nm UV illumination intensity from 9.7 mWmm\textsuperscript{-2} to 4.9 mWmm\textsuperscript{-2} has little effect on the speeds of the microswimmers, indicating that already at 4.9 mWmm\textsuperscript{-2} the illumination is almost saturating. A decrease in the instantaneous velocities is observed upon reducing the H\textsubscript{2}O\textsubscript{2} concentration by 4x, in line with previous reports on the relationship between fuel concentration and microswimmer speeds \cite{Howse2007}. We note that the relative decrease in the instantaneous velocities is small compared to the reduction in H\textsubscript{2}O\textsubscript{2}, indicating that low fuel concentrations are viable for motion experiments. The most evident decrease in microrobot speeds is observed when increasing the wavelength of UV illumination from 340 nm to 380 nm (decreasing the incident photon energy). The band gap of TiO\textsubscript{2} lies between 360 and 380 nm, which likely explains the more significant reduction in instantaneous velocity observed. For the analysis of photo-responsive behavior and orientation-dependent velocities, we use the experimental conditions providing the highest median velocity (340 nm illumination, 9.4 mWmm\textsuperscript{-2}, 3v\% H\textsubscript{2}O\textsubscript{2} concentration).

\newpage
\setcounter{figure}{0}
\renewcommand{\figurename}{Fig. S}
\renewcommand{\tablename}{Table S}

\section*{Supplementary Figures}

\begin{figure}[h]
\centering
   \includegraphics[width=0.95\linewidth]{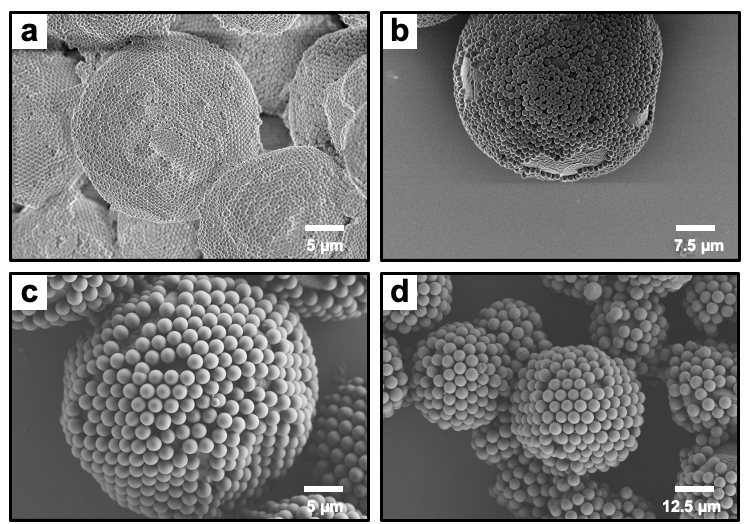}   
  \caption{Colloidosomes obtained with: a. 0.55 $\mu$m particles b. 1.2 $\mu$m particles c. 2.12 $\mu$m particles d. 4.16 $\mu$m particles }
   \label{suppfig:1}
\end{figure}

\begin{figure}
\centering
   \includegraphics[width=0.95\linewidth]{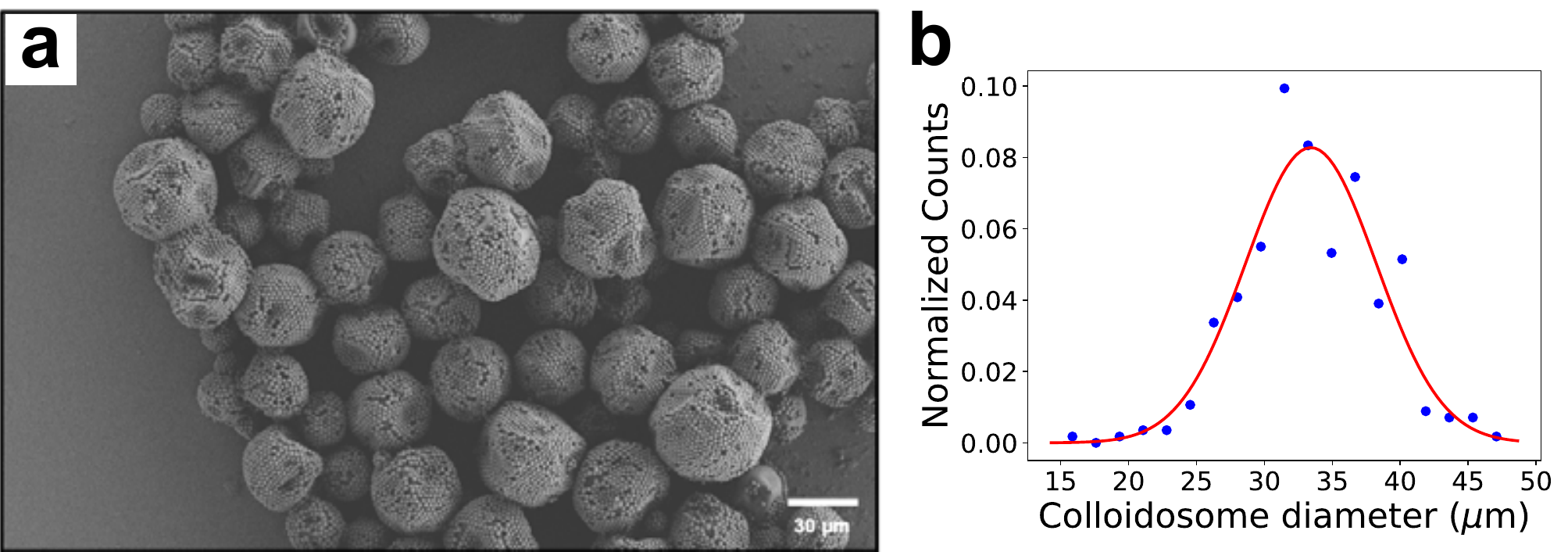}  
  \caption{a. Colloidosomes obtained from 2.1 $\mu$m particles b. Size distribution of pictured colloidosomes; from a sample size of 325, we observe a Gaussian distribution centred at 33.42 $\mu$m with a standard deviation of 4.82 $\mu$m}
   \label{suppfig:2}
\end{figure}

\begin{figure}
    \centering
   \includegraphics[width=0.95\linewidth]{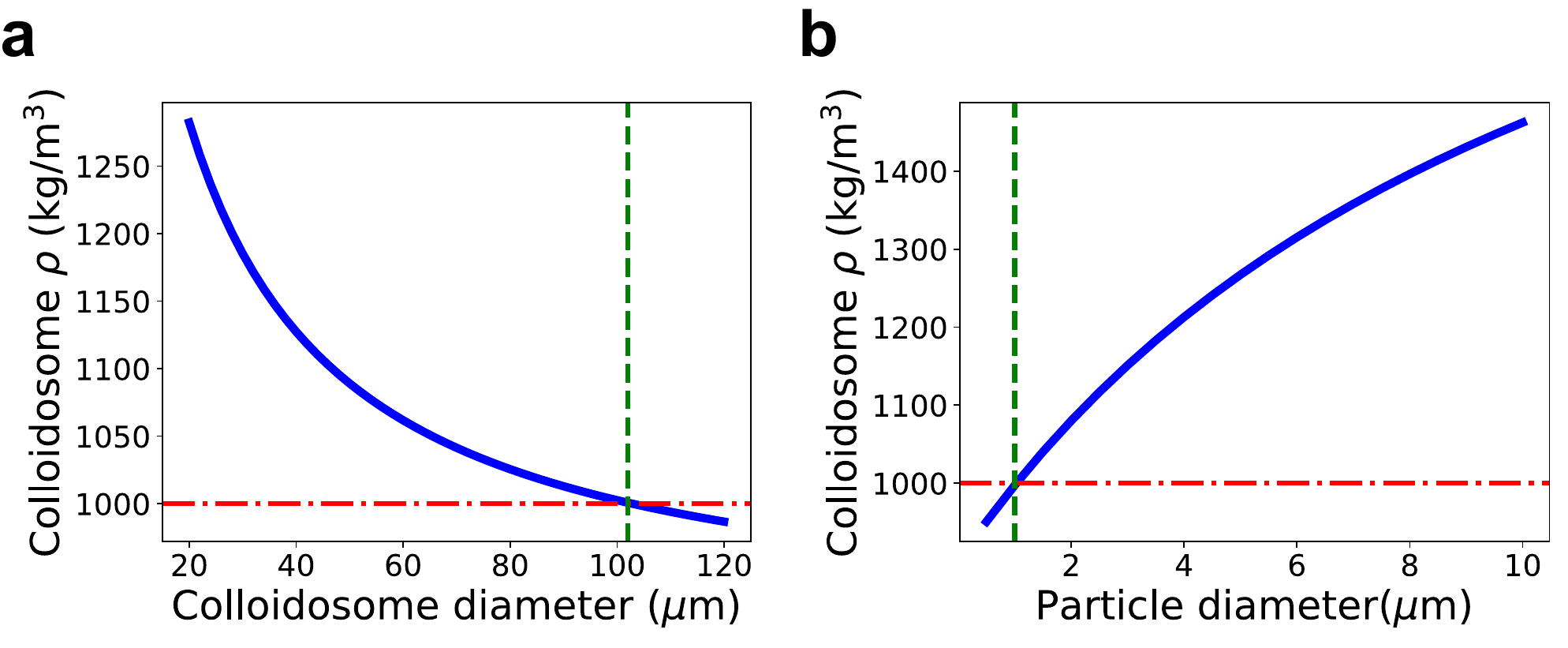} 
  \caption{a. Colloidosome (from 2 $\mu$m SiO\textsubscript{2} particles) density as a function of its size. b. Density of a 50 $\mu$m colloidosome with increasing stabilizing SiO\textsubscript{2} particle size. The red dashed line marks the density of water, and the green dashed line marks the intersection of colloidosome density and water}
   \label{suppfig:3}
\end{figure}

\begin{figure}
  \centering
   \includegraphics[width=0.8\linewidth]{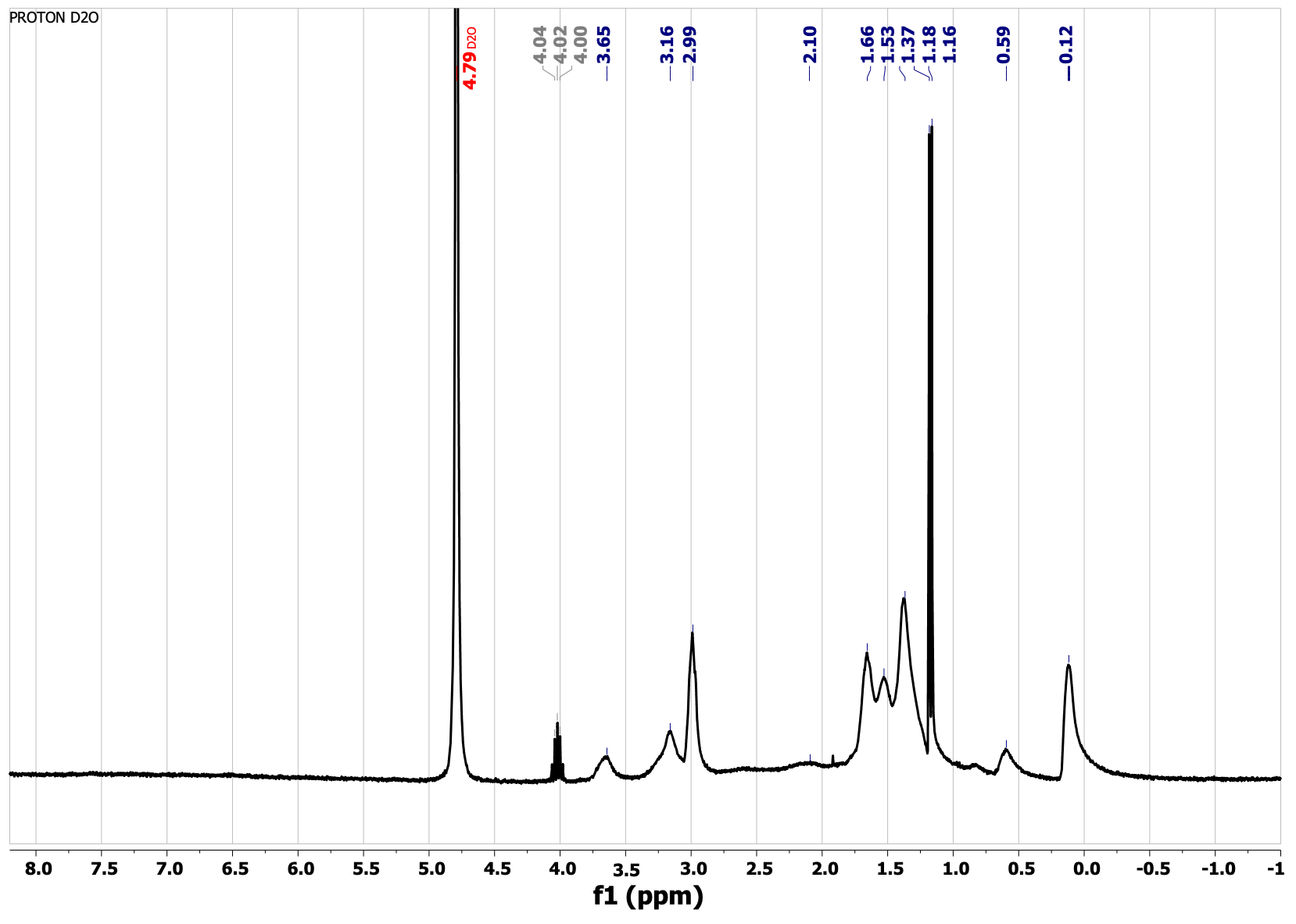} 
  \caption{D\textsubscript{2}O \textsuperscript{1}HNMR spectrum of post-modified poly(acrylamide)-g-(1,6-hexanediamine, 3-aminopropyl-dimethyloxysilane, nitrodopamine)}
   \label{suppfig:4}
\end{figure}

\begin{figure}
  \centering
  \includegraphics[width=0.8\linewidth]{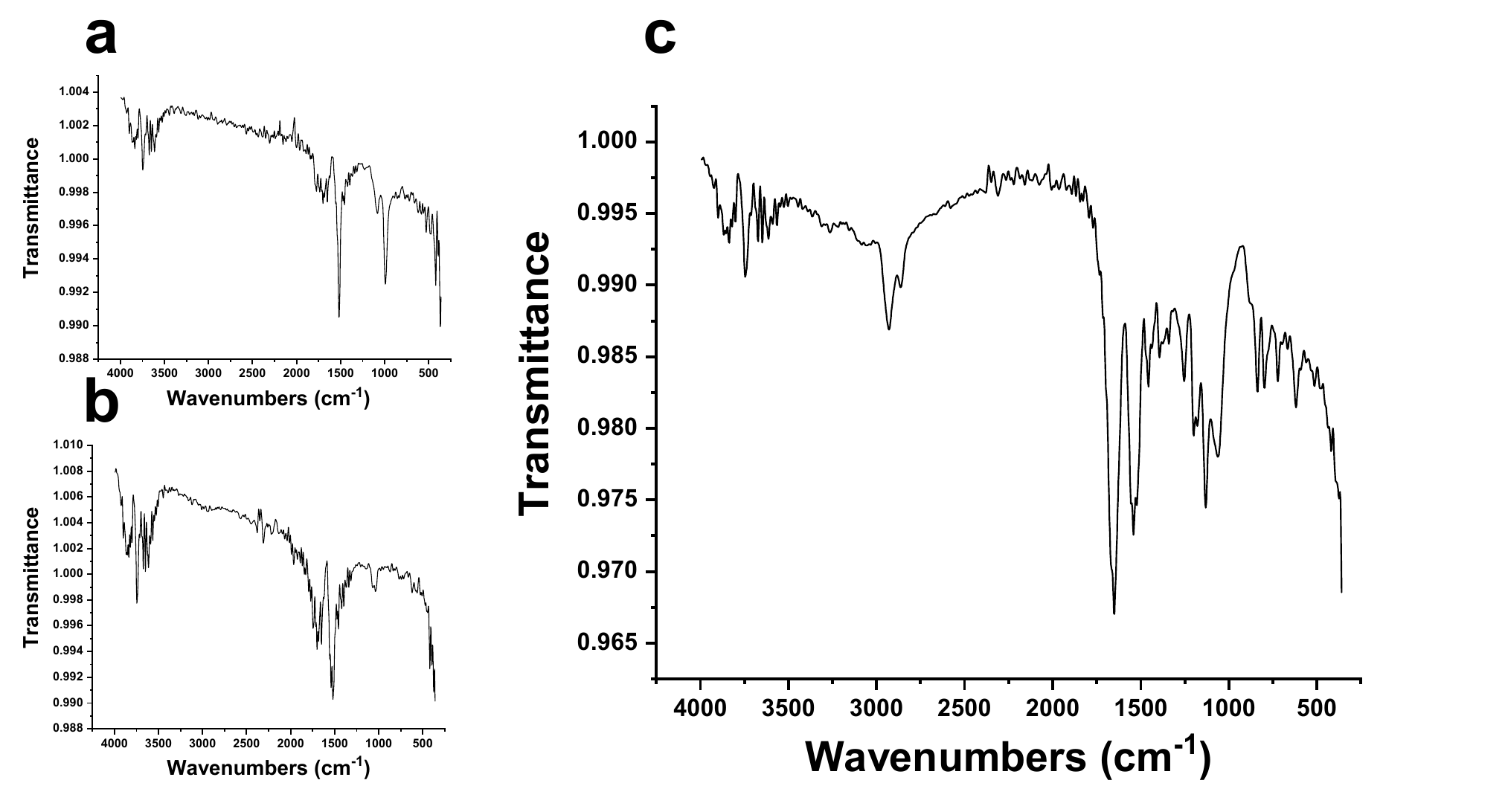} 
  \caption{ATR-IR spectra of a. Unmodified pPFPAC backbone b. Nitrodopamine c.  Post-modified poly(acrylamide)-g-(1,6-hexanediamine, 3-aminopropyl-dimethyloxysilane, nitrodopamine) }
   \label{suppfig:5}
\end{figure}

\begin{figure}
   \includegraphics[width=\linewidth]{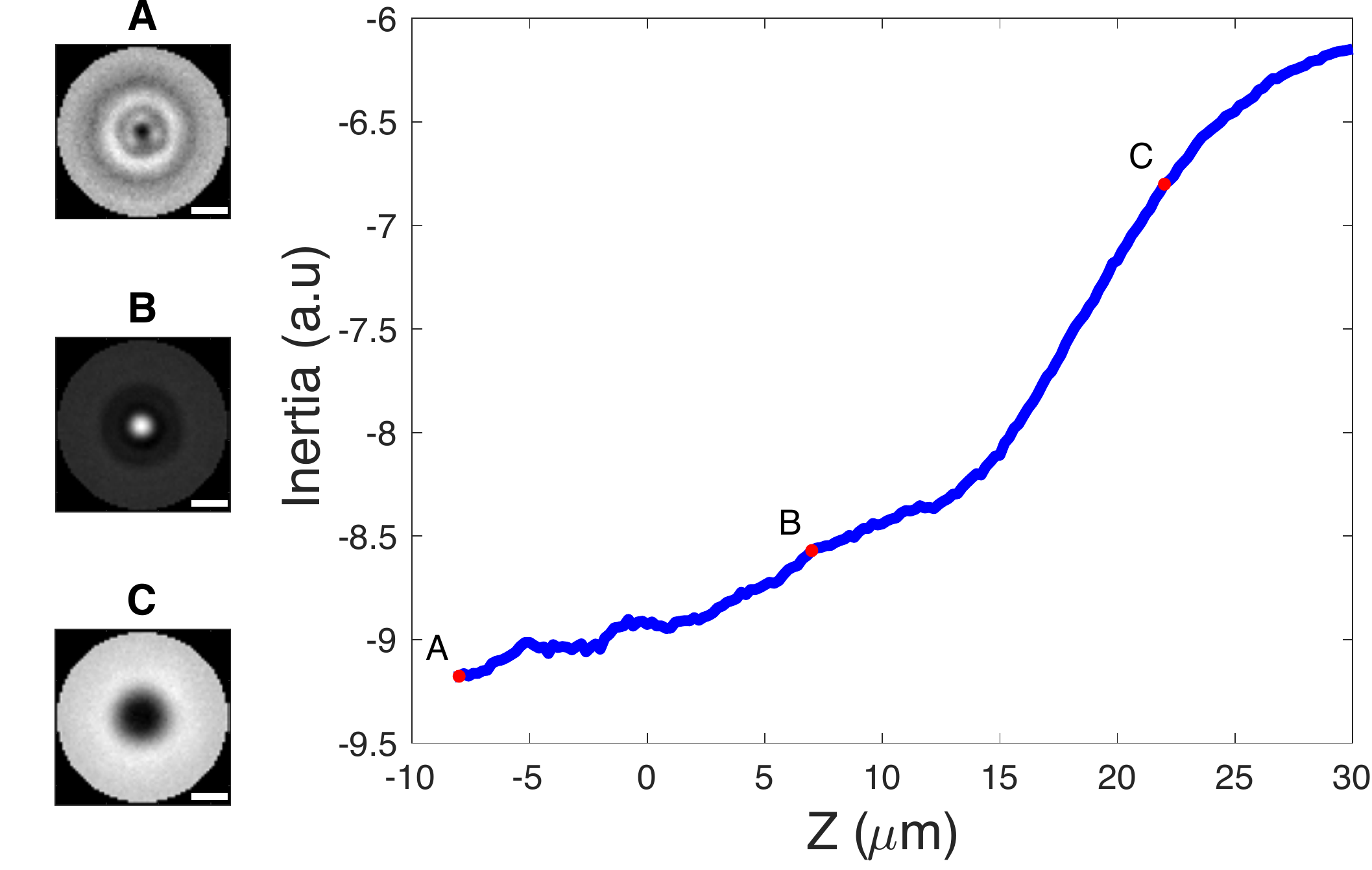} 
  \caption{LUT produced of Inertia(Z) from Z-stacks of immobilised particles. The monotonic region of interest is studied. The LUT is inverted to obtain Z(Inertia), to which a cubic polynomial is fit by nonlinear regression. Left: Images of particles taken at different Z heights. Right: Inertia(Z) LUT averaged from 5 Z-stacks, with the particles on the left indicated. Scale bars represent 2$\mu$m }
   \label{suppfig:6}
\end{figure}

\begin{figure}
  \centering
  \includegraphics[width=0.8\linewidth]{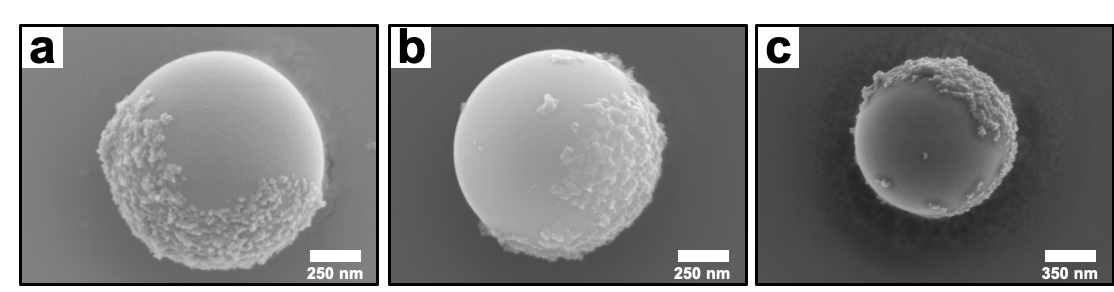} 
  \caption{a-c: Microrobots after exposure to highly oxidative conditions (long UV exposure in an H\textsubscript{2}O\textsubscript{2}-rich solution). The morphology of the nanoparticle thin films remains unchanged.}
  \label{suppfig:7}
\end{figure}

\begin{figure}
  \centering
  \includegraphics[width=0.8\linewidth]{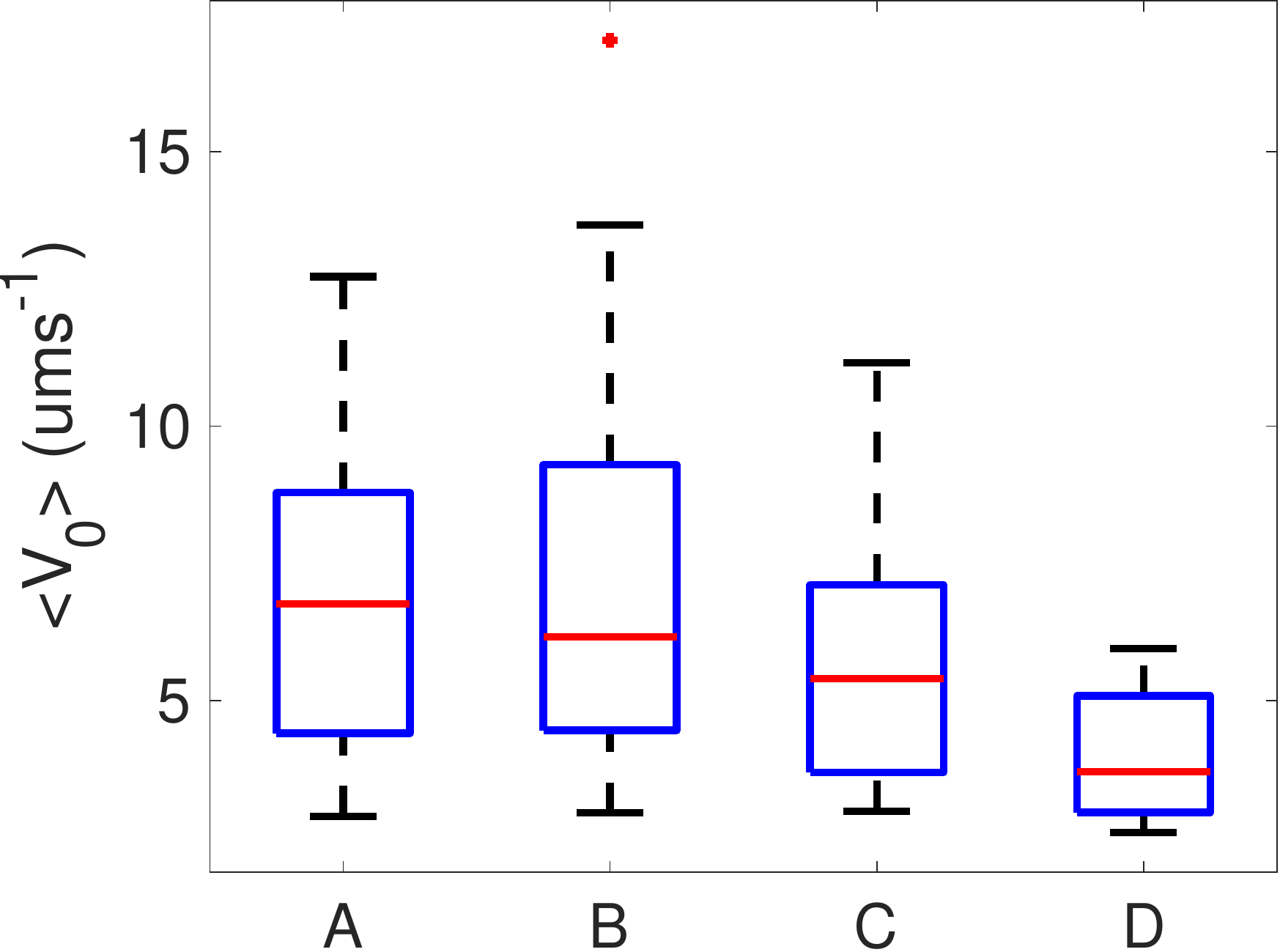} 
  \caption{Box-plots of instantaneous velocities (evaluated over 0.2s) obtained under different conditions as follows: A) 340nm 9.4mW/mm\textsuperscript{2}, H\textsubscript{2}O\textsubscript{2} 3v\% B) 340nm 4.9mW/mm\textsuperscript{2}, H\textsubscript{2}O\textsubscript{2} 3v\% C) 340nm 9.4mW/mm\textsuperscript{2}, H\textsubscript{2}O\textsubscript{2} 0.75v\% D) 380nm 61.7mW/mm\textsuperscript{2}, H\textsubscript{2}O\textsubscript{2} 3v\% }
  \label{suppfig:8}
\end{figure}

\newpage
\section*{Supplementary Tables}
\begin{table}[h]
  \centering
  \includegraphics[width=0.95\linewidth]{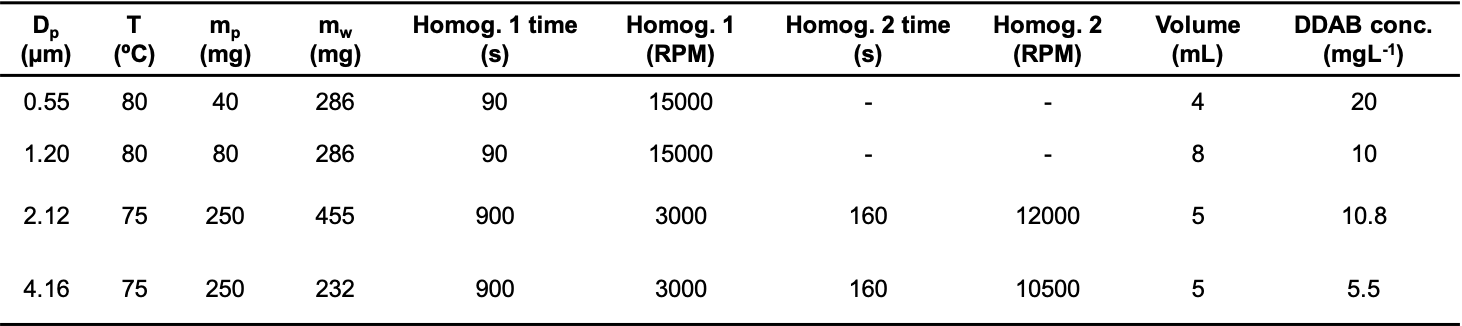}
  \caption{Conditions used to obtain colloidosomes with particles of different sizes}
  \label{supptbl:1}
\end{table}

\begin{table}[h]
  \centering
  \includegraphics[width=0.95\linewidth]{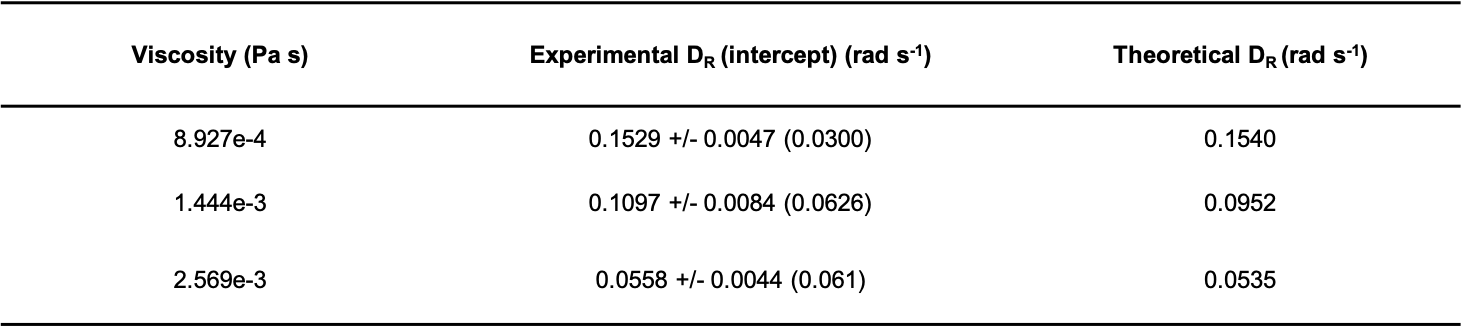} 
  \caption{Rotational diffusion for single particles in different viscosity solutions found following the method proposed by Anthony et al \cite{Anthony2006}. The theoretical values expected from the Stokes-Einstein relationship ($D\textsubscript{R} = k\textsubscript{B}T/8\pi \eta R\textsuperscript{3})$ are also shown as reference.}
  \label{supptbl:2}
\end{table}

\end{document}